          [2001/04/25 1.1 (PWD)]
\documentclass[a4paper,twocolumn]{esapub2005} 
\usepackage{graphicx}
\usepackage{times}
\usepackage{natbib}

\newcommand{\sax}{{\it BeppoSAX} }
\newcommand{\asca}{{\it ASCA} }

\newcommand{\exosat}{{\it EXOSAT} }
\newcommand{\einstein}{{\it Einstein} }
\newcommand{\integral}{{\it INTEGRAL} }

\newcommand{\temna}{{\it Temna} }

\newcommand{\source}{4U~1636--53 }

\begin{document}
\title{Hard tail detection in the LMXB \source from \integral\/
}
\author{Mariateresa  Fiocchi}
\author{Angela Bazzano}
\author{Pietro Ubertini}
\author{Memmo Federici}
\affil{Istituto di Astrofisica Spaziale e Fisica Cosmica di Roma (INAF). Via Fosso del Cavaliere 100, Roma, I-00133, Italy}
\maketitle
\begin{abstract}
Recent \integral\/ observations and archival \sax\/ data 
have been analyzed to deeply investigate
the hard X-ray behavior of the 
neutron star,
atoll type, low mass X-ray
Binary \source.
Our investigation in three different periods outline three
corrisponding different sates.
Infact, the source was detected in both the canonical high and low state and 
moreover in one occasion 
\integral spectrum shows, for first time in this source,
a hard tail dominating the emission above 30 keV.
This spectrum is fitted as the sum of
a Comptonized component similar to soft state
and a power-law component ($\Gamma=2.76$), indicating the presence of a
non thermal electron distribution of velocities.
\end{abstract}
\section{Introduction}
\source is a neutron star low mass X-ray binary (LMXB)
classified as a atoll source \citep{has89},
with an orbital period of 3.8 hr derived from the optical
variability of its companion V801 Arae \citep{ped84}
and at distance of 3.7--6.5 kpc \citep{fuj88, sma92, aug98}.
While the X-ray burst properties and timing signatures have been analyzed extensively
(see \citep{jon05}, \citep{bel05} and references therein)
the spectral characteristics have been studied only at low energy with
\einstein\/ \citep{chr97}, \exosat\/ \citep{vac87}, \temna\/
\citep{wak84} and \asca\/ \citep{asa98}.
In general, the spectrum
was acceptably fitted by a Comptonization model plus a black body component.
We report here a broad band spectral analysis
performed on data from \sax and \integral satellites, which allowed us to better
constrain the spectral parameters and to detect the presence of a high energy tail
dominating the spectrum above $\sim 30$ keV.
A similar feature has been observed in other LMXBs namely GX~17+2
\citep{dis00},
GX~349+2 \citep{dis01},
Sco~X-1 \citep{dam01},
4U~1608-522 \citep{zha96}, XB~1254-690 \citep{iar01},
Cir~X-1 \citep{iar02} and 4U~0614+091 \citep{pir99}.
\section{Observations and Data Analysis}
Table \ref{jou}  summarizes the log of \integral and \sax observations of \source.
\sax observed the source on three occasions: February and March 1998 and
February 2000.
LECS, MECS and PDS event files and spectra,
available from the ASI Scientific Data Center,
were generated by means of the Supervised Standard Science Analysis
\citep{fio99}.
Both LECS and MECS spectra were accumulated in circular regions
of 8' radius.
The PDS spectra were obtained
with the background rejection method based on fixed rise time thresholds.
Publicly available matrices were used for all the instruments.
Fits are performed in the following energy band: 0.5--3.5 keV for LECS, 1.5--10.0
keV for MECS and 15--70 keV for PDS.\\
The analyzed \integral \citep{win03} data set consists of all observations in which
4U~1636-53 was
within the high-energy detectors field of view.
Observations are organized into uninterrupted 2000 s long science pointing, windows
(scw):
light curves and spectra are extracted for each individual scw.
Wideband spectra (from 5 to 150 keV) of the source are obtained using data from the
two high-energy
instruments JEM-X \citep{lun03} and IBIS \citep{ube03}.
Data were processed using the Off-line Scientific Analysis
(OSA version 5.1)
software released by the \integral Scientific Data Centre.
Data from the Fully Coded field of view only for both instrument have been used.
\section{Spectral Analysis}
\label{analisis}
Several physical models have been used to fit the
whole
data set,
keeping the number of free parameters as low as possible.
Each time a new component was added to the model, a F-test was performed. We
assumed that a F probability
larger than $95\%$ implies a significative improvement of the fit.
\begin{table*}
\centering
\caption{BeppoSAX and INTEGRAL Observations}
\label{jou}
\scriptsize
\begin{tabular}{lccccccc}
\hline\hline
\multicolumn{8}{c}{BeppoSAX Journal}\\
& {\bf Start Date}&&{\bf Exposure time}&&&{\bf Count s$^{-1}$}&\\
& & & & & & &\\
&&LECS&MECS&PDS&LECS&MECS&PDS\\
&&ksec&ksec&ksec&[0.4-3 keV]&[1.5-10 keV]&[20-60 keV]\\
& & & & & & &\\
\emph{$1^{st}$ epoch (a)} &1998-02-24 &13 &39 &17 &$14.66\pm0.03$  &$39.37\pm0.02$$^{a}$ &$0.48\pm0.04$\\
\emph{$1^{st}$ epoch (b)} &1998-03-01 &6  &14 &7  &$14.95\pm0.05$  &$25.42\pm0.04$$^{b}$ &$0.80\pm0.06$\\
\emph{$1^{st}$ epoch (c)} &2000-02-15 &12 &37 &19 &$22.08\pm0.04$  &$29.09\pm0.03$$^{b}$ &$0.73\pm0.03$\\
\hline
\multicolumn{8}{c}{INTEGRAL Journal}\\
& & & & & & &\\
&&{\bf Start Date}&\multicolumn{2}{r}{\bf Exposure time}&\multicolumn{2}{c}{\bf Count s$^{-1}$}&\\
&&&JEM-X&IBIS&JEM-X&IBIS&\\
&&&ksec&ksec&[5-15keV]&[20-150 keV]&\\
& & & & & &&\\
&\emph{$2^{nd}$ epoch} &2003-03-04  &36          &594         &$6.3\pm0.2$   &$10.74\pm0.08$&\\
&\emph{$3^{rd}$ epoch} &2003-03-04  &16          &117         &$3.70\pm0.06$ &$3.22\pm0.04$&\\
\hline
\hline
\end{tabular}\\
{
$^{a}$ MECS count rates refer to MECS2 and MECS3 units.
$^{b}$ MECS count rates refer to only MECS2 unit.
}
\end{table*}
The uncertainties are at $90\%$ confidence level for one
parameter of interest ($\Delta\chi^2=2.71$).\\
When spectra are from more than one detector,
we allow the relative normalization to be free with respect to the
MECS and IBIS data, for \sax\/ and \integral\/ respectively.
XSPEC v.\ 11.3.1. has been used.\\
Spectral behavior has been studied separately in three epochs consisting of the following data:
\begin{itemize}
\item
\emph{$1^{st}$ epoch}: all three \sax observations available from February 1998
to February 2000.
During these periods the source was always in a soft/high state.
\item
\emph{$2^{nd}$ epoch}: JEM-X and IBIS data available from 52644 MJD to
53644 MJD with count rate $<5~counts~s^{-1}$ in the 20--40 keV energy band. For the chosen period
the source was in a hard/low state.
\item
\emph{$3^{rd}$ epoch}: JEM-X and IBIS data available from 52644 MJD to 53644 MJD
with count rate $>5~counts~s^{-1}$ in the 20--40 keV energy band.
This epoch does not corrspond to either the soft or the hard state and
and here we call it
\textit{peculiar} state as will be explained in detail later on.
\end{itemize}
After veryfing there were no changes in the three observations and in order 
to
achieve the highest signal to noise ratio,
we build the \sax average spectrum arranging
all three observations.
We then can
take advantage of the high quality of the average spectrum up to about 70 keV.
The most simple model which provides a good fit to the average \sax spectrum
in the energy band 0.5--70 keV
consist of a thermal Comptonized component modeled in XSPEC by
{\scriptsize{COMPTT}}
(a spherical geometry was assumed) plus
a soft component which we modeled with two temperature blackbody.
Being the column density {N$_{\rm H}$} towards the source
measured by the LECS and MECS instrument
always close to the galactic column density
($N_{\rm H}~galactic~=~3.58\times10^{21}~cm^{-2}$,
estimated from the 21 cm measurement \citep{dic90}),
we fix this parameter at this value.
Spectral fit results are given in Table \ref{fit}, the average spectrum is shown in
Fig.\ 1.

\begin{small}
\begin{figure}
\label{sax}
\centering
\includegraphics[angle=-90,width=7.0cm]{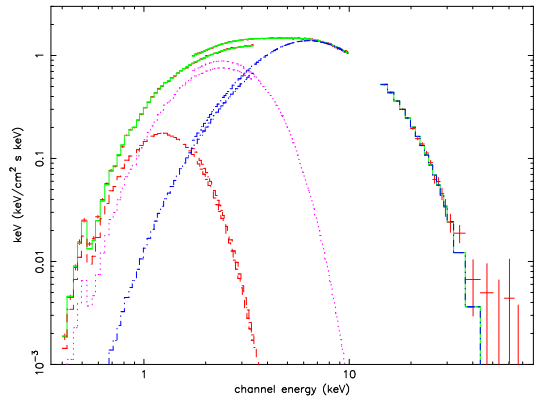}
\caption{The \sax average spectrum in the soft state (epochs 1), shown together with
the total model and its components: the total, two blackbody and the Comptonization
components are shown in green, red, magenta and blue, respectively.
}
\end{figure}
\end{small}
The source was in the soft/high state
with an un-absorbed luminosity of $L_{0.1-200\,\rm keV} \simeq 2.0 \times 10^{37}$ 
erg s$^{-1}$, assuming a distance of 5.9 kpc (Cornelisse et al. 2003).
As in the case of the \einstein observation \citep{chr97},
thermal Comptonization of the
optically thick plasma corona with a quite low electron temperature
is dominating.\\
The most simple model which provides a good fit to the \integral\/ hard state
consist of a thermal Comptonized component modeled in XSPEC by
{\scriptsize{COMPTT}} ~\citep{tit94}
(a spherical geometry was assumed) plus
a soft component which we modeled by a single temperature blackbody.
Because of the very good low energy \sax coverage, we used
the average value of input soft photon temperature and column density
($T_0=1.3 keV$ and $N_H=N_H~Galactic $) measured by the
BeppoSAX observations.
During this period, the source has a
luminosity of $L_{0.1-200\,\rm keV} \simeq 1.4 \times 10^{37}$ erg s$^{-1}$,
lower than the one in the soft state,
in agreement with the usual ranking of the luminosity
in atoll sources
(e.g., \citep{has89}; \citep{van00}; \citep{gie02}).
 The electron temperature is now substantially higher than in the soft state,
$T_{\rm e}\sim 23$ keV, and the Comptonization component extends well above $\sim 100$ keV.
The same model was been applied to the
$2^{nd}$ \integral\/ data set but resulted in a poor fit with
a $\chi^2/d.o.f=97/57$ with clear residuals above 60 keV.
Adding a power law improves the fit significantly
($\chi^2/d.o.f$
becomes $64/55$).
The disk component becomes negligible and it is not necessary to best fit the data.
Spectral fit
 results are given in Table \ref{fit}, and spectra are shown in Fig.\ \ref{int1}.
Since the spectral parameters in the \textit{peculiar} state are similar
to those in the soft/high \sax state, a possible interpretation of this state is
that \source was in a similar soft/high state
as during the \sax observations,
but with a new overlapped component at high energies simultaneously present.
The un-absorbed luminosity is $L_{0.1-200\,\rm keV} \simeq 7.4 \times 10^{37}$ erg s$^{-1}$.\\
\begin{table*}
\scriptsize
\caption{Results of the fit of \source spectra in the energy band $0.5-70$ keV and $5-150$ keV
for \sax and \integral observations respectively, 
for three different spectral states.}
\label{fit}
\centering
\begin{tabular}{ccccccccc}
\hline\hline
\multicolumn{9}{c}{\bf{BeppoSAX average high/soft state spectrum}}\\
\multicolumn{9}{c}{$model=bbody+bbody+comptt$}\\
{T$_{\rm BB1}$} & {T$_{\rm BB2}$}& $T_{0}$  &{T$_{\rm e}$} &$\tau$ & $R_{\rm BB1}$& $R_{\rm BB2}$& $n_{\rm COMPTT}$ & $\chi^2$/d.o.f \\
$keV$ &$keV$ &$keV$& $keV$ &&km &km& &\\
& & & & & & &\\
$0.24\pm0.01$&$0.58\pm0.02$&$1.3\pm0.1$ &$3.4\pm0.3$ &$3.8\pm0.4$ &$79\pm6$ &$25\pm1$ &$0.19\pm0.02$&235/190\\
\hline
\multicolumn{9}{c}{\bf{INTEGRAL low/hard state spectrum}}\\
\multicolumn{9}{c}{$model=bbody+comptt$}\\
&{T$_{\rm BB2}$}& $T_{0}$  &{T$_{\rm e}$} &$\tau$ & $R_{\rm BB2}$ & $n_{\rm COMPTT}$ & $\chi^2$/d.o.f& \\
&$keV$ &$keV$& $keV$ &&km &$10^{-2}$ &&\\
& & & & & & &\\
&$1.2_{-0.3}^{+0.2}$&1.3~fixed &$23_{-2}^{+7}$&$1.1_{-0.3}^{+0.5}$&$5^{+3}_{-2}$ &$1.0^{+0.3}_{-0.1}$ &57/59&\\
\hline
\multicolumn{9}{c}{\bf{INTEGRAL peculiar state spectrum}}\\
\multicolumn{9}{c}{$model=comptt+powerlaw$}\\
&$T_{0}$  &{T$_{\rm e}$} &$\tau$ &$\Gamma$ & $n_{\rm COMPTT}$ & $n_{\rm powerlaw}$&$\chi^2$/d.o.f& \\
&$keV$&$keV$& &&$10^{-2}$ &$ph~keV^{-1}~cm^{-2}$&&\\
& & & & & & & &\\
&1.3~fixed&$5^{+4}_{-3}$ &$3\pm2$&$2.6\pm0.1$&$2.8_{-0.6}^{+0.8}$ &$1.0^{+1.4}_{-0.5}$ &72/57&\\
\hline
\hline
\end{tabular}\\
\end{table*}
\begin{small}
\begin{figure}
\centering
\includegraphics[angle=-90,width=7.0cm]{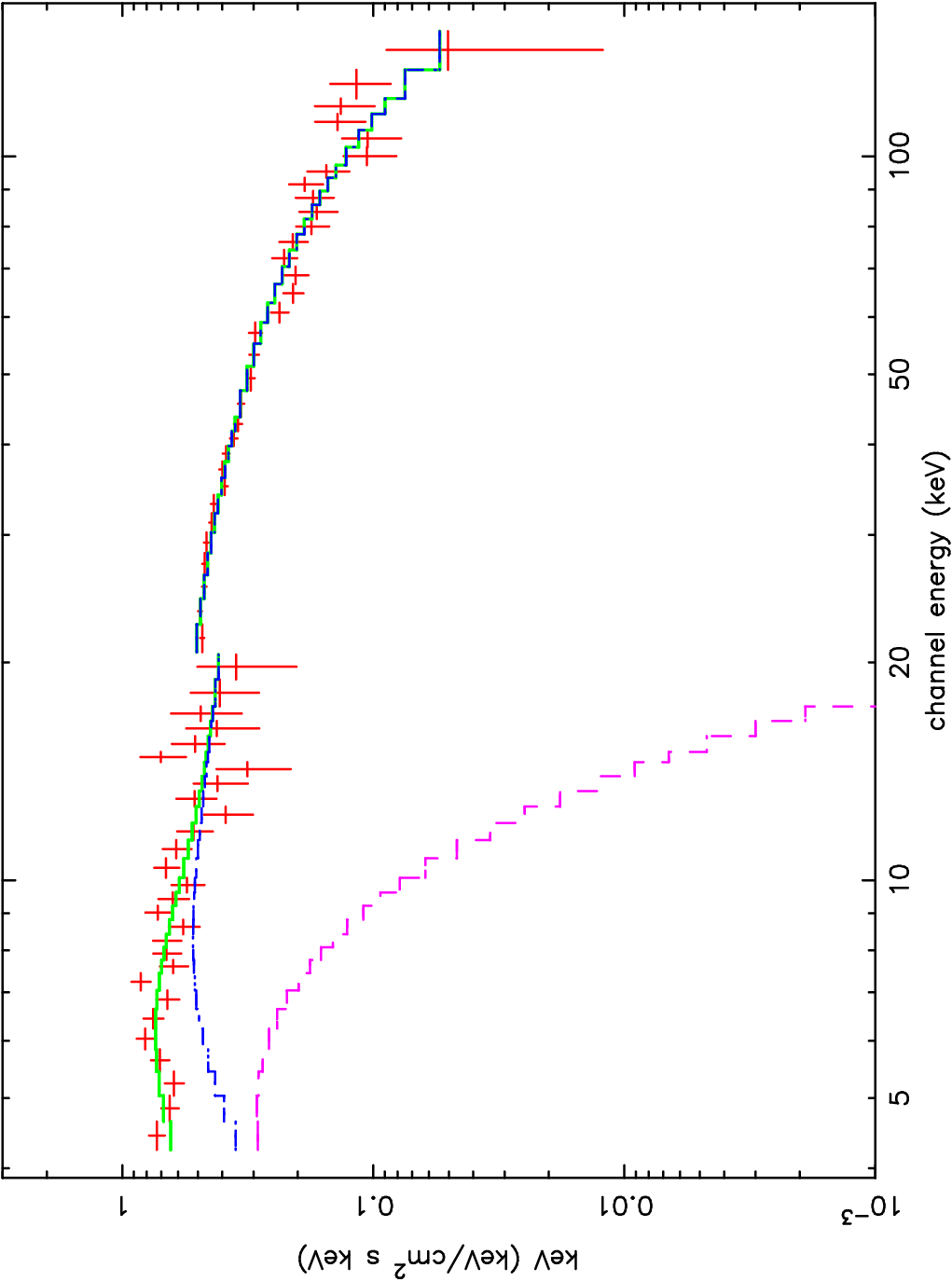}
\includegraphics[angle=-90,width=7.0cm]{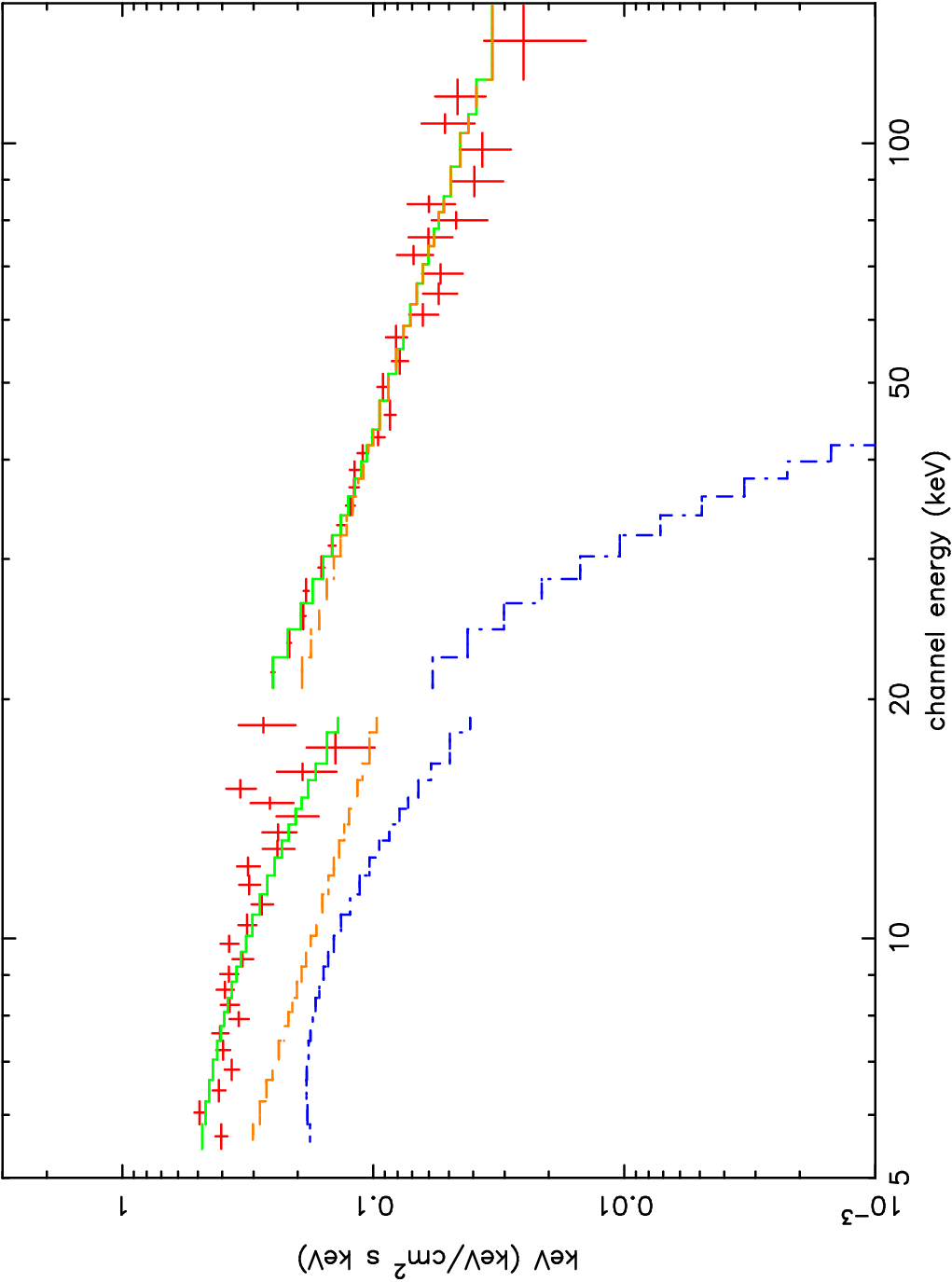}
\caption{The spectra of epochs 2 (soft state, left) and 3 (\textit{peculiar} state, right)
observed by \integral, shown together with the total model and its components. Left: the total,
the blackbody and the Comptonization components are shown in green, magenta and blue, respectively.
Right: the total, the Comptonization and the power law component are shown in green,
blue and orange respectively.}
\label{int1}
\end{figure}
\end{small}
\section{Conclusions}
\label{discussion}
The black-body component in the soft states could originate at both the
neutron star surface and the surface of an optically-thick accretion disk.
In our observations the two black body components seems to originate
from two different parts of the disk,
corrisponding to two different temperature.
The Comptonization component may arise from a corona above the disk and/or
between the disk and the neutron star surface.
In the hard states, accretion probably assumes the form of a truncated
outer accretion disk 
as previously reported
by \citep{bar02} for LMXB 4U~1705-44.
The spectral transitions are generally, but not necessarily,
coupled with changes in luminosity,
indicating they are driven by variability of the accretion rate
or change of the geometry of the system as for Black Hole hard/soft transition
at constant luminosity \citep{bel05}.
For \source the accretion rate is lower in the hard state than in the soft state
and becames very high in the {\it{peculiar}} state. 
We note 
this value can be influenced by the steep power law
that dominates the
energy spectrum at low energies.
The emission is best described
by Comptonization from a complex
electron distribution due to a low temperature
($\sim 5$keV) thermal electron distribution
together with
non-thermal power-law electrons.
This two-component electron distribution
could be explained by the following hypothesis:
\begin{itemize}
\item
Non-thermal electron
acceleration regions powered by magnetic reconnections above a
disc. Low-energy electrons cool preferentially by Coulomb
collisions leading to a thermal distribution while the
high-energy electrons cool by Compton scattering, preserving a
non-thermal distribution \citep{cop99}.
\item
The
magnetic reconnection above the disc can produce a
non-thermal electron distribution,  while overheating of the
inner disc produces the thermal Comptonization \citep{kub01}.
\item
The power-law component could be produced by Componization by
synchrotron emission in the jet \citep{bos05, fen04}
In our source, this hypothesis is strengthened by radio detection:
Sydney University Molonglo Sky Survey catalog gives a flux of 7.5 mJy at 843 GHz
\citep{mau03}.
\end{itemize}
Finally, Laurent \& Titarchuk \citep{lau99} suggest the detection of a power-law
component at high energy
to be a signature of presence black hole in an X-ray binary system. Our data
are supporting result on GX~17+2 \citep{dis00},
clearly showing this criterion proposed to distinguish black hole versus
neutron star binaries is inadequate.
\begin{small}

\end{small}
\section*{Acknowledgments}
We acknowledge the ASI financial/programmatic support via contracts ASI-IR
046/04.
A special thank to K. Kretschmer for making data
available before becoming public.

\end{document}